\documentclass[12pt,preprint]{aastex}

\shorttitle{Afterglow of GRB\,021004}
\shortauthors{Laursen \& Stanek}

\begin{document}

\title{High-Precision Photometry of the Gamma-Ray Burst 020813: The
Smoothest Afterglow Yet}

\author{L.~T.~Laursen, K.~Z.~Stanek}

\affil{Harvard-Smithsonian Center for Astrophysics, 60
Garden St., Cambridge, MA 02138}

\email{llaursen@cfa.harvard.edu, kstanek@cfa.harvard.edu}

\begin{abstract}

We report results of our precise reduction of the high signal-to-noise
$VRI$ observations of the optical afterglow of the gamma-ray burst
GRB\,020813 obtained by Gladders \& Hall with the Magellan 6.5-m
telescope $3.9-4.9\;$hours after the burst. These observations are
very well fitted locally by a power-law curve, providing the tightest
constraints yet on how smooth the afterglows can be in some cases: the
rms deviations range from $0.005\;$mag (0.5\%) for the $R$-band to
$0.007\;$mag for the $I$-band, only marginally larger than the rms
scatter for nearby non-variable stars.  This scatter is a factor of
several smaller then the smallest reported rms of $0.02\;$mag for
GRB\,990510 (Stanek et al). These observations are in strong contrast
to those of afterglows of GRB\,011211 and GRB\,021004, for which large
$> 10$\% variability has been observed on timescales from $\sim 20$
minutes to several hours.  If interstellar medium (ISM) density
fluctuations near the GRB are indeed causing the bumps and wiggles
observed in some bursts, very uniform regions of ISM near some bursts
must be present as well. This result also constrains the intrinsic
smoothness of the afterglow itself.

\end{abstract}


\section{Introduction}
\label{sect_intro}

Short-timescale variability in the optical afterglow of gamma-ray
bursts (GRBs) can be a tool for understanding the details of GRB
origins. The optical afterglow of a GRB stems from a relativistic
blastwave as it slows down in the interstellar medium/stellar wind
surrounding a source hypernova (e.g. Stanek et al. 2003; Matheson et
al. 2003). Understanding variations in intensity of the optical
afterglow, or lack thereof, can yield insights into the details of the
interaction between GRBs and their surroundings (e.g. Wang \& Loeb
1999). Such $> 10$\% variability has been observed by Holland et
al. (2002) in GRB\,011211 and by Bersier et al. (2003) in GRB\,021004.

GRB\,020813 was detected by the HETE2 at 2:44:19 UT on 2002 August 13
(Villasenor et al. 2002). Its optical afterglow was localized by
\citet{gcn1470} at $\alpha_{2000} = 19^h 46^m 41\fs 88,\ \delta_{2000}
= -19\arcdeg 36\arcmin 05\arcsec$. The properties of the burst and the
afterglow have been described so far by Barth et al. (2003), Covino et
al. (2003), Li et al. (2003) and Urata et al. (2003).  Gladders \&
Hall (2002a,b) began taking optical data $3.0\;$hours after the event
with the Baade 6.5-m telescope.  These data had exceptionally good
seeing and high signal-to-noise, and were made public by the authors
via an {\tt anonymous ftp}. We decided to use these high-quality data
to investigate the possible presence of short-timescale variability in
the afterglow. In this paper we find that between $3.9-4.9\;$hours
after the burst the afterglow of GRB\,020813 has been the smoothest
yet, with rms deviations ranging from $0.005\;$mag for the $R$-band to
$0.007\;$mag for the $I$-band, only marginally larger than the
photometric scatter for nearby non-variable stars.

\section{Observations and Data Reduction}

\label{sect_phot}

The $VRI$ data were obtained with the Las Campanas Observatory
Magellan Baade 6.5-m telescope equipped with the TEK5 camera by
M. Gladders and P. Hall over two nights (Gladders \& Hall 2003c).
Thirty-seven 60 s exposures were taken of the afterglow the first
night (11 in $V$, 10 in $R$, and 16 in $I$), and fifteen of the same
length the second night (2 in $V$, 4 each in $BR$, and 5 in
$I$). Since we were interested in the short-term variability, we
decided not to reduce the second night's data.  According to the {\tt
ftp} posting\footnote{
ftp://ftp.ociw.edu/pub/gladders/GRB/GRB020813/README}, they were
overscan corrected, trimmed, de-biased, and flat fielded using
calibration frames from the first night (August 13 UT). Also, a fringe
frame was produced from other science observations on August 13 UT and
was used to de-fringe the $I$-band data.

We used the DAOPHOT point-spread function fitting package (Stetson
1987, 1992) and the ISIS image subtraction package (Alard \& Lupton
1998, Alard 2000) to reduce the data. We found excellent agreement
between the two packages.  For consistency, we used the photometry
obtained with DAOPHOT throughout this paper. Images from the first
night were brought to a common zero point using approximately 50 stars
per image, providing very stable differential photometry.  Our
reduction of the Gladders \& Hall data is listed in
Table~\ref{phottable}.

While this is not important for the current paper, for consistency we
have calibrated our photometry to that of Covino et al. (2003).  We
notice that the reductions of Gladders \& Hall's $R$-band data by
Covino et al. (2003) and by Li et al. (2003) differ by about
$0.09\;$mag, with the Li et al. photometry being brighter.

\vspace{-0.3cm}
\section{Short Timescale Variations}
\label{sec_Short term variations}

\begin{figure}
\plotone{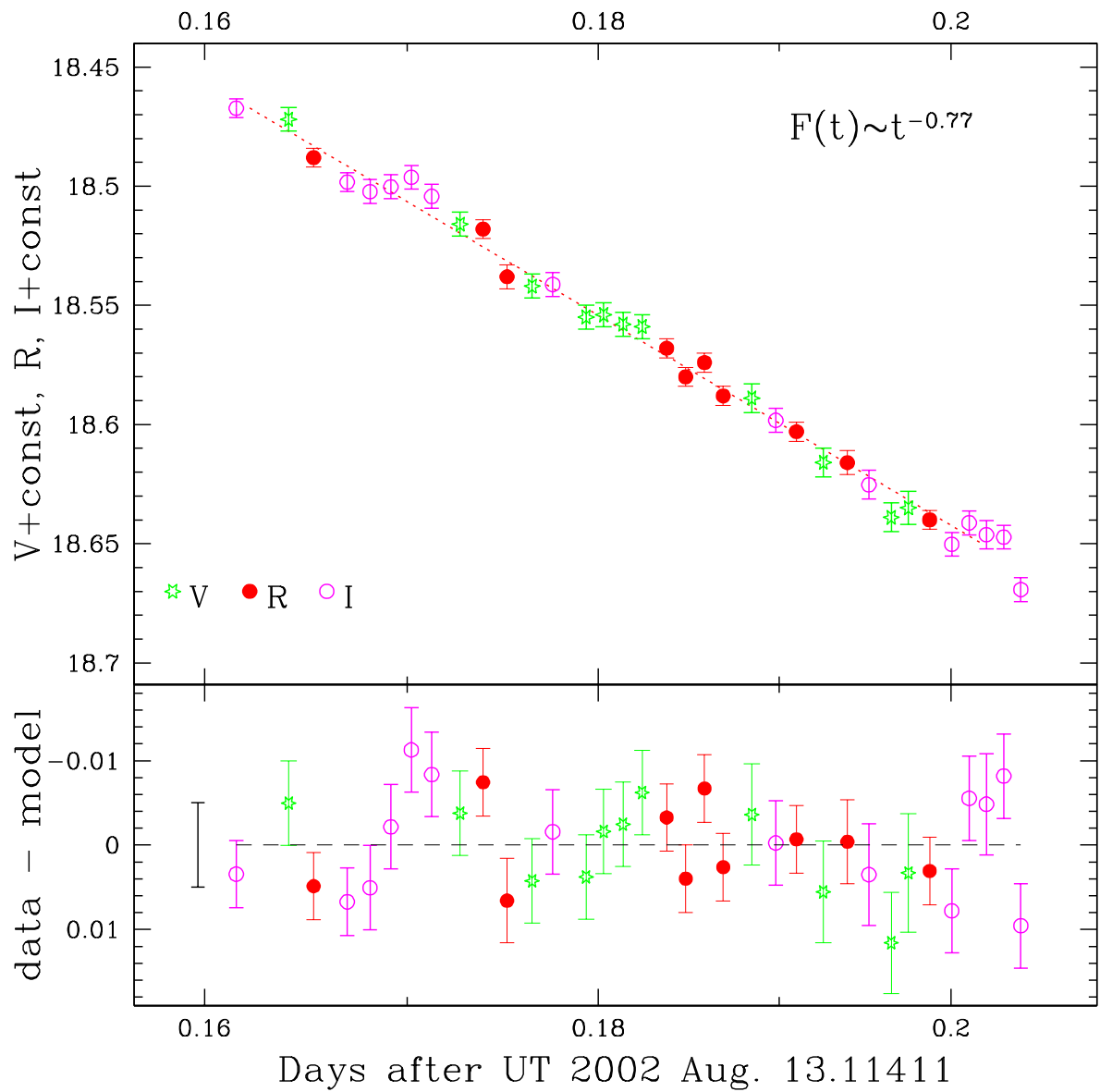}
\caption{{\it Upper panel}\/ $VRI$ light curve of the optical
afterglow of GRB\,020813 during $3.9-4.9\;$hours after the burst. A
power law has been fitted to the data.  The $V$ and $I$ data have been
shifted.  {\it Lower panel}\/ Residuals between the power law model
and the $VRI$-band data. The error bar on the left is typical for
non-variable stars with brightness similar to the afterglow ($rms \sim
0.005$ mag in $I$-band).
\label{fig_fit}}
\end{figure}

We fitted a power-law to first night $VRI$ data; this yielded a local
decay slope of $0.77$. This simple fit turns out to be a good
description of the OT temporal behavior. In the $R$-band, the
residuals from the power-law behavior are the smallest, with rms of
$0.005\;$mag. The rms is slightly larger, $0.006\;$mag, for the
$V$-band, and is the largest for the $I$-band, $0.007\;$mag.  For
comparison, nearby non-variable stars show rms scatter of $\sim
0.003\;$mag in the $R$-band, $\sim 0.005\;$mag in the $I$-band and
$\sim 0.007\;$mag in the $V$-band. The observed deviations might be
marginally significant for the $RI$-bands. However, these deviations
are a factor of several smaller then the smallest deviations reported
so far: for GRB\,990510, $R$-band rms scatter of $\sim 0.021\;$mag was
observed, with the largest deviation from the smooth decay being
$0.08\;$mag (Stanek et al. 1999; see also Hjorth et al. 1999).

So far there have been two detections of short timescale variations in
optical afterglows: GRB\,011211 (Holland et al. 2002) and, especially
clear, GRB\,021004 (Bersier et al. 2003).  Models have attempted to
explain short timescale bumps and wiggles in optical GRB light curves,
but given the smoothness of the curve presented here, consideration
should be given to explaining very smooth curves as well. If ISM
density fluctuations near the GRB explain the bumps and wiggles, the
models must also allow for very uniform regions of ISM.

The data reduced here covers only an hour very early after the event,
hence it cannot provide limits for later interaction with the ISM or
other behavior. Longer-term, equally dense and high quality sampling
would be appropriate for a better understanding of short-timescale
variability over the course of the GRBs. Such data most likely already
exist, for example for GRB\,030329, which was very bright and observed
intensively by numerous observers for many days. While the light curve
of GRB\,030329 was very bumpy on timescales of days, on timescales of
hours the light curve was very smooth (e.g. Matheson et al. 2003).

\vspace{-0.3cm}
\acknowledgments

We thank M. Gladders and P. Hall for making their data publicly
available, which made this project possible. LTL thanks A. Bonanos for
her unfailing assistance during the data reduction procedure. We thank
D. Bersier, S. T. Holland and S. Jha for their comments on an earlier
version of this paper. LTL was supported by a Harvard College Research
Program Fellowship.

\begin{deluxetable}{ r c c c }
\tabletypesize{\footnotesize}
\tablewidth{0pt}
\tablecaption{MAGELLAN PHOTOMETRY\label{phottable}}
\tablehead{\colhead{$\Delta$T\tablenotemark{a}} &
\colhead{Mag} &
\colhead{$\sigma_m$} &
\colhead{Filter} }
\startdata
0.1641 & 18.903 & 0.005 & V \\
0.1727 & 18.947 & 0.005 & V \\
0.1765 & 18.973 & 0.005 & V \\
0.1793 & 18.986 & 0.005 & V \\
0.1803 & 18.985 & 0.005 & V \\
0.1813 & 18.989 & 0.005 & V \\
0.1824 & 18.990 & 0.005 & V \\
0.1884 & 19.020 & 0.006 & V \\
0.1925 & 19.047 & 0.006 & V \\
0.1965 & 19.070 & 0.006 & V \\
0.1975 & 19.066 & 0.007 & V \\
0.1653 & 18.488 & 0.004 & R \\
0.1739 & 18.518 & 0.004 & R \\
0.1752 & 18.538 & 0.005 & R \\
0.1837 & 18.568 & 0.004 & R \\
0.1848 & 18.580 & 0.004 & R \\
0.1858 & 18.574 & 0.004 & R \\
0.1868 & 18.588 & 0.004 & R \\
0.1910 & 18.603 & 0.004 & R \\
0.1939 & 18.616 & 0.005 & R \\
0.1987 & 18.640 & 0.004 & R \\
0.1615 & 17.939 & 0.004 & I \\
0.1670 & 17.970 & 0.004 & I \\
0.1681 & 17.974 & 0.005 & I \\
0.1692 & 17.972 & 0.005 & I \\
0.1702 & 17.968 & 0.005 & I \\
0.1713 & 17.976 & 0.005 & I \\
0.1776 & 18.013 & 0.005 & I \\
0.1898 & 18.070 & 0.005 & I \\
0.1952 & 18.097 & 0.006 & I \\
0.2001 & 18.122 & 0.005 & I \\
0.2011 & 18.113 & 0.005 & I \\
0.2021 & 18.118 & 0.006 & I \\
0.2032 & 18.119 & 0.005 & I \\
0.2042 & 18.141 & 0.005 & I
\enddata 
\tablecomments{[The complete version of this table is in the
electronic edition of the Journal. The printed edition contains only a
sample.]}  
\tablenotetext{a}{Days after 2002 August 13.11411 UT.}
\end{deluxetable}

\end{document}